\documentclass[]{llncs}
\usepackage[T1]{fontenc}
\usepackage{graphicx}

\usepackage{preamble}
\usepackage{macros}

\begin{document}

\title{A Formally Verified Lightning Network}

% LNCS
\author{Grzegorz Fabiański\inst{1} \and Rafał Stefański\inst{1} \and Orfeas Stefanos Thyfronitis Litos\inst{2}}
\institute{University of Warsaw \and Imperial College London}

% ACMART
%\author{Grzegorz Fabiański}
%\affiliation{%
%  \institution{University of Warsaw}
%  \country{Poland}
%}
%\email{grzegorz.fabianski@mimuw.edu.pl}
%
%\author{Rafał Stefański}
%\affiliation{%
%  \institution{University of Warsaw}
%  \country{Poland}
%}
%\email{rafal.stefanski@mimuw.edu.pl}
%
%\author{Orfeas Stefanos Thyfronitis Litos}
%\affiliation{%
%  \institution{Imperial College London}
%  \country{United Kingdom}
%}
%\email{o.thyfronitis-litos@imperial.ac.uk}

%\author{First Author\inst{1}\orcidID{0000-1111-2222-3333} \and
%Second Author\inst{2,3}\orcidID{1111-2222-3333-4444} \and
%Third Author\inst{3}\orcidID{2222--3333-4444-5555}}

\maketitle
\begin{abstract}
In this work we use formal verification to prove that the Lightning Network (LN), the most
prominent scaling technique for Bitcoin, always safeguards the funds of honest
users. We provide a custom implementation of (a simplification of) LN, express
the desired security goals and, for the first time, we provide a 
machine checkable proof that they are upheld under
every scenario, all in an integrated fashion.
We build our system using the Why3 platform.
%\keywords{First keyword  \and Second keyword \and Another keyword.}
\end{abstract}

\section{Introduction}
Bitcoin~\cite{bitcoin} is the oldest and consistently the highest valued
block\-cha\-in. However, despite its value and prominence,
it faces severe scalability issues~\cite{scaling} both in
terms of throughput and of latency: it natively supports only up to $7$
transactions per second and requires $1$ hour from transaction submission to its
verification.

To alleviate this limitation, the \emph{Lightning Network} (LN)~\cite{lightning}
has been developed. It lifts payments \emph{off-chain}, i.e., it enables parties
to securely pay each other without interacting with the block\-chain. LN users
enjoy latency and throughput limited only by the communication network between
the transacting parties, avoid the hefty per-payment on-chain fees, and
relieve other block\-chain users from having to reach consensus on every
single payment.

As of this writing, over \$$120$M are contained in LN. This makes LN a
security-critical application. Nevertheless, to the best of the authors'
knowledge, its complex
design\footnote{\url{https://github.com/lightning/bolts/blob/master/00-introduction.md}}
has never been formally verified for crucial properties such as safeguarding
honest parties' funds. The current work aims to fill this gap. We employ the
Why3~\cite{boogie11why3} formal verification platform to implement a subset
of LN, formally specify the desired security properties and mechanically prove
that our implementation adheres to them under all execution paths.
Our system consists of $6060$ lines of program code and $7420$ lines of
verification code.

\subsection{Overview of LN}
The central construct of LN is the \emph{payment channel}. Each channel has
exactly two parties, each of which has some coins in it. One party can pay the
other via an update of the channel state, which is done collaboratively with its
counterparty. Each party is guaranteed that they can
\emph{unilaterally} close the channel on-chain and get their fair share of
coins at any time. Channel updates are very cheap: the counterparties just need
to exchange a
few messages --- no interaction with third parties or the block\-chain is needed
for paying within one's channel.

In slightly more detail, LN works as follows: To initially transfer coins from
the block\-chain to a new channel, the two parties collaboratively create a
``joint account''. This involves the two parties generating three Bitcoin
transactions, with one submitted on-chain (the \emph{funding transaction}) and
the other two stored locally off-chain (the \emph{commitment transactions}). The
funding transaction transfers the initial coins to the joint account, whereas
the commitment transactions safeguard each party's access to its coins: each
party can publish on-chain one of the commitment transactions to move its coins
back to the block\-chain without needing the cooperation of its counterparty. This
ensures that no trust between parties is needed. All in all, when published
on-chain, the funding transaction opens the channel and the commitment
transaction closes it.

On each payment, two things happen: Firstly a new pair of commitment transactions
is generated, reflecting the balance of coins after the payment. Secondly, the
previous pair of commitment transactions are \emph{revoked}. A revoked
commitment transaction corresponds to an old channel state and thus should never
be published on-chain. A malicious party can nevertheless publish it --- this
could be beneficial if the older balance were in its favor. In such a
case, the honest counterparty can publish the corresponding \emph{revocation}
transaction (which was generated when the offending commitment transaction was
revoked) within a timeframe and punish its counterparty by confiscating all
channel coins.

To sum up, a channel needs one on-chain transaction to open, one on-chain
transaction to close and can support a practically unlimited number of off-chain
payments. It offers low latency, high throughput, no fees and needs no extra trust
requirements compared to the underlying block\-chain.

% TODO to shorten: remove next paragraph (apart from last sentence, which can go
% to \ref{subsec:flow}
The main focus of this work is the security of a single, two-party channel.
Channel operations consist of funding a new channel, processing off-chain payments,
and closing the channel in a way that the rightful
funds are returned to the two parties on-chain. We do not model
networks of multiple parties, nor the multi-hop payments supported by production
implementations.

\subsection{Why3}
We implement and specify LN in Why3~\cite{boogie11why3}, a framework for
formally verifying high-level code. Why3 focuses on proof automation using SMT
solvers. It is largely composed of two parts: Firstly, there is the
WhyML language, used to express programs. It also supports expressing
formal specifications in first-order logic in the form of inline assertions,
separate lemmas, or function pre- and
post-conditions; this allows for natural code constructs. Secondly, there is the \emph{driver} which translates the verification conditions of
WhyML code into SMT queries. The novelty of Why3 lies in the relatively seamless
integration of a variety of SMT solvers. In the current work we have taken
advantage of this feature by using both Alt-Ergo~\cite{conchon:hal-01960203} and
CVC4~\cite{DBLP:conf/cav/BarrettCDHJKRT11}.

Another important formal verification tool commonly used in the context of
cryptographic protocols is EasyCrypt~\cite{10.1007/978-3-642-22792-9_5}. We note that we chose not to formalize LN using
EasyCrypt because the latter is more geared towards low-level, probabilistic
reasoning and computational assumptions ---
features that are used by most cryptographic protocols, but challenging for the
general-purpose formal verification tools. In contrast, our approach assumes idealized low-level
primitives (i.e., digital signatures) and proves security unconditionally
(i.e., without a negligible probability of failure), therefore is largely
incompatible with EasyCrypt. In fact, our approach of
deterministically modeling signatures obviates the need for EasyCrypt, thereby
simplifying our overall proof effort.

\subsection{Preliminaries on Cryptography} \label{subsec:preliminaries-on-cryptography}
LN relies on \emph{digital signatures}, which consist of
three algorithms: key generation, signing and verification. In our work, we
adopt the \emph{ideal functionality} defined in~\cite{1310743} to model these operations.
Instead of running the algorithms, parties send queries to and receive outputs
from the functionality, which internally keeps track of the signed messages and
embodies the single source of truth, obviating the need for randomness
and hard-to-model computational assumptions (see
Subsec.~\ref{subsec:signatures}).
Since the functionality masks the need for such assumptions, we can
strengthen our Why3 modeling of the adversary to an unbounded one. In fact, instead of modeling the
adversary as a polynomially bounded Turing machine in the context of Why3, the
security result proven by our code is against any sequence of choices
made by the adversary, even non-computable ones.
Although strengthening the adversary in this way
might seem counterintuitive, it actually makes the verification effort simpler and clearer.
The powers of such an adversary are directly expressed via \emph{universal non-determinism},
which is readily available in Why3. We note that our high-level result
(Def.~\ref{def:funds-sec}) directly carries over to bounded adversaries.
Therefore our definition also covers real-world systems which, due to their use
of practical constructions of the digital signature algorithms, can only be
secure against bounded adversaries.

\subsection{Our Contributions}
This work consists of the following contributions:
\begin{itemize}
\item We provide an implementation of the two-party LN interactions in
Why3,
\item We implement the subset of Bitcoin logic that is relevant to LN in Why3,
\item We define \emph{funds security} for an honest client in the presence of a
Byzantine counterparty,
\item We provide a machine-checkable proof of the funds security of LN channels,
\item We demonstrate how to define \emph{funds ownership} in the presence of
multisig accounts and signed transactions stored off-chain,
\item We identify crucial properties that any LN implementation should uphold to
guarantee funds security and use them to modularize our proof. This
provides a blueprint for proving the security of production LN implementations.
\end{itemize}

\subsubsection{Limitations.}
%(i.e., no HTLCs, no multi-hop payments, no LN/on-chain fees).
The main simplifications in our implementation compared to the real Lightning Network (LN)
is that it only supports direct payments between two parties. As such it does not support 
multi-hop payments or HTLCs. Furthermore, we use a simplified model of Bitcoin which,
in particular, does not include on-chain fees (see Section~\ref{sec:bitcoin} for more details).

\subsection{Protocol Flow in Our Implementation}
\label{subsec:flow}
Our work focuses on a single, two-party LN channel over Bitcoin. We implement a
fragment of the LN functionality, namely direct payments between two parties. We
do not include HTLCs and multi-hop payments in our implementation.

The operation of our channel closely follows that of production LN
implementations. When one of the two parties (a.k.a. the \emph{funder}) is
instructed to open a new channel, the funding process begins. The funder
notifies its counterparty (a.k.a. the \emph{fundee}) and the two parties engage
in a series of messages, at the end of which the channel is established. All
initial funds belong to the funder and they originate from the funder's on-chain
bitcoins.
%If any of the two parties fails to respond appropriately at any point,
%then its counterparty will not lose any money.

Once a channel has been established, a payment can be initiated by either side. The payer specifies the amount of
in-channel funds to transfer, which is subtracted from the payer's channel
balance. After a few messages, the payee's balance is increased by the payment
sum. Channels support an arbitrary number of sequential payments.

At any moment, either party can initiate the closing procedure. An honest party
also closes its channel if its counterparty misbehaves (e.g., stops responding).
An honest closing party is guaranteed to receive its channel balance on-chain
after bounded delay. Internally, channel closure is initiated by submitting a
single transaction on-chain that reflects the channel balance. This starts a
timer, during which the counterparty can dispute the claimed channel balance
if the closing party submitted an outdated channel state. Honest client operation guarantees
that the honest party loses no funds due to failed disputes.

\subsection{Security Goals}
We consider two standard properties of LN:
The first one, which we call \emph{funds security}, states that a party
can always close the channel and transfer its rightful funds on-chain within a bounded 
time frame. For the sake of funds security, we assume that the honest party interacts with a malicious,
byzantine party that also controls the environment.
The second property called \emph{liveness} states that channel funding
and channel payments between two honest parties will always complete in a bounded amount of time.
For the sake of liveness, we assume that two parties are honest, but the adversary controls the network 
and payment requests (see Subsec.~\ref{subsec:adversarial-model} for more details).

In this work, we only prove funds security, not liveness.
In fact, we are aware that in a scenario when the two honest parties try to initiate payments
simultaneously in both directions, our implementation might deadlock. (This does not violate 
the guarantee of funds security, as the deadlock can be resolved by either party closing the channel).
Nonetheless, we do test that a simple channel opening and payment scenario completes successfully
(see \texttt{twoHonestParties.mlw}). 
As the violation of funds security is far more detrimental than the violation of liveness,
we believe that such an approach strikes a reasonable balance between proof complexity
and practical relevance.

\subsection{Adversarial Model}\label{subsec:adversarial-model}
In our model, funds security is formalized via a $1$-player game, following
established cryptographic practice. The game is played by the adversary.
Different moves in the game correspond to different actions that the adversary
can carry out in real life. For example, it can choose to deliver a message to
the ledger, or to have the corrupted channel party send arbitrary messages to
its honest counterparty. These two moves correspond to the power to control the
network and to corrupt one channel party respectively. Its control over the
corrupted party is complete: it can send and sign arbitrary messages on its
behalf at any point in time. On the other hand, its control over the network is
encumbered with the responsibility to deliver transactions to the ledger within
a specific time bound. Furthermore, it controls the moments in which the honest
party is activated, but it must ensure that the delay between two activations
does not exceed a specific time bound. 
These limitations correspond to the standard, reasonable network assumptions and
the security of LN depends on them.

Last but not least, we give the adversary the power to choose the moment of
channel opening and closure, as well as the timings, directions and amounts of
payments while the channel is open. These additional adversarial powers
correspond to the choices that the honest party can make in a practical
deployment. We choose this approach as it models the worst-case scenario, making
our results stronger as well as conformant with the standard approach in
cryptography. Allowing these choices to be made by the adversary
has the added benefit of ensuring that the game is strictly $1$-player, keeping
the model simpler.
The adversary wins if a channel closure request is put forward but, after the
aforementioned time bound, the honest party has not received its rightful funds
on-chain.

\subsection{Architecture Overview}
\label{subsec:architecture-overview}
We now discuss our code architecture at a high level --- Appx.~\ref{sec:components} contains a relevant diagram. We provide a simplified
implementation of Bitcoin called $\Gamma$, which focuses on the capabilities
related to LN (Sec.~\ref{sec:bitcoin} and \texttt{gamma.mlw}\footnote{The full
code can be found at
\url{https://github.com/grzegorzFabianski/LightningNetworkInWhy3} --- \texttt{README.md} contains
relevant documentation.}). Furthermore, we implement the signature
functionality~\cite{1310743}. It is used by the honest party, the adversary, and
$\Gamma$ to sign messages and verify signatures (Subsec.~\ref{subsec:preliminaries-on-cryptography}
and \texttt{signaturesFunctionality.mlw}).
Our LN client can read the state of $\Gamma$ and
send transactions to it via a network queue controlled by the adversary. The
honest client receives instructions to perform actions, such as to open a
channel or carry out a payment. These instructions come from the adversary, but
invalid instructions are ignored by the honest client (\texttt{honestPartyType.mlw} and
\texttt{honestPartyInteractions.mlw}). Last but not least, we
formally define funds security, as discussed in Sec.~\ref{sec:model}
(\texttt{honestPartyVsAdversary.mlw}).

In order to keep the proof tidy in the face of the complex interactions between
the aforementioned components, we use a number of techniques
(Sec.~\ref{sec:proof}). We here highlight the two most salient ones. The LN
client discussed above has to keep track of a complex state which includes many
low-level details, such as which messages remain to be sent in order to conclude an
in-flight payment. In order to keep these details separate from the proof, we
identify a simple set of requirements on the client behavior
(see Subsec.~\ref{subsec:party-interface}, Appx.~\ref{app:party-interface}, and \texttt{honestPartyInterface.mlw}).
We can thus separate our proof
into two independent parts: We first prove that our LN implementation meets
these requirements and we then prove that any implementation that satisfies
these requirements enjoys funds security. We believe that our requirements must
be satisfied by any secure LN implementation, therefore a proof of security of
an existing production LN deployment boils down to just proving that it
conforms to our requirements.

The second proof technique elucidates the meaning of coin ownership. Albeit a
useful high-level concept, is not straightforward to define due to signed but
non-processed transactions and multisig accounts. To that end, we define the
\texttt{Evaluator} (Subsec.~\ref{subsec:ownership}, Appx.~\ref{sec:evaluator}, and module \texttt{Evaluator} in \texttt{gamma.mlw})
which formally answers the
question ``How much money does a party own?'' This abstraction greatly
simplifies the proof effort. We believe this approach can be reused in similar
verification projects.

\subsection{Related Work}

\subsubsection{Formal verification of blockchain infrastructure and
applications.}
Existing blockchain-related formal verification efforts revolve around two axes:
Verifying consensus protocols and verifying smart contracts.

With respect to consensus protocols, HotStuff~\cite{10.1145/3293611.3331591} has
been formally specified and verified by~\cite{10.1007/978-3-030-77448-6_9} using
TLA+~\cite{lamport1999specifying}. Furthermore, parts of the
Tendermint~\cite{tendermint} consensus reference implementation have been
formally verified in~\cite{10.1007/978-3-030-61362-4_27} using TLA+. Similarly
to both of these protocol analyses, the current work expresses the execution of
multiple parties that communicate via the network, the delays of which are
explicitly modeled. Moreover, similarly to~\cite{10.1007/978-3-030-77448-6_9}
and~\cite{10.1007/978-3-030-61362-4_27}, we verify end-to-end guarantees of the
execution, not just static invariants.

Regarding the verification of smart contracts, a survey of
tools for Ethereum smart contract analysis, including formal verification tools,
can be found in~\cite{8782988}. Concrete examples are
Manticore~\cite{8952204}, EthVer~\cite{10.1007/978-3-662-63958-0_30},
PRISM~\cite{10.1007/978-3-642-22110-1_47},
and~\cite{electronics11193091,10.1007/978-3-030-53288-8_8,hildenbrandt-saxena-zhu-rodrigues-daian-guth-moore-zhang-park-rosu-2018-csf}.
These tools focus on specific contract
properties, whereas our work takes a whole-system approach, whereby the smart
contract logic (i.e., the LN transactions) only form one part of the entire
system, the other parts being the explicit modeling of parties, the
communication network they use, and the process of signing transactions.

\subsubsection{Tools and formal verification projects using Why3.}
\label{subsubsec:why3-tools}
A formal verification framework that has seen extensive use in a variety of
major projects in the area is Why3~\cite{boogie11why3}. Some of these
projects aim at providing higher-level, special-purpose verification tools.
Michelson, the low-level language of smart contracts for the Tezos
block\-chain~\cite{9188227}, can be automatically translated into WhyML
with~\cite{9284726}. 

\subsubsection{Security analyses of LN.}
A survey of results on LN security and attacks against it can
be found in~\cite{10224687}. LN security has been formally modeled and proven
in the  UC setting in~\cite{9155145}. Most comparable to
our work is~\cite{grundmann2023formal}, which builds upon~\cite{9805487} in an effort to
formally verify the security of LN using TLA+.

Both~\cite{9155145} and~\cite{grundmann2023formal} model the intricate details
of funds security in the multi-hop payment setting, while our work focuses on a
single channel. Due to the lack of multi-hop payments, the wormhole and griefing
attacks~\cite{10224687} do not apply in our work. With respect to funds
security, the main difference between our work
and~\cite{9155145,grundmann2023formal} is the framework used. Any small
differences in the exact security guarantees result from framework
differences, not from high-level security goals.

We now present a detailed comparison of our work with~\cite{grundmann2023formal}, starting with the similarities.
Both works prove security for a specification of LN, as opposed
to a specific implementation. Both approaches model signatures as ideal functionalities, removing their
inherent randomness (the latter is problematic in the context of formal verification)
in the process. Last but not least, in both cases non-determinism is used to
model the adversary, including its exact leeway for arbitrary behavior.

On the other hand, the current work differs from~\cite{grundmann2023formal} in
the following respects: We model security using a game-based definition, proving
specific security properties, whereas~\cite{grundmann2023formal} uses a
simplified version of UC-based modeling. We use
the Why3 first-order proof system, which allows us to model infinite system
states, whereas~\cite{grundmann2023formal} uses the TLA+ model checker and only
checks all possible scenarios up to a specific number of actions. Our
modeling is limited to simple payments within a single channel,
whereas~\cite{grundmann2023formal} further models multi-hop payments over
multiple channels. We provide a custom implementation of LN,
upon which we base our specification of LN, which happens to be more abstract
than the official specification (BOLT), whereas~\cite{grundmann2023formal}
verifies the BOLT specification itself. Last but not least, our formal
verification effort is complete, whereas~\cite{grundmann2023formal} constitutes
an intermediate report of a proof effort that is still underway.
\section{Security Model}
\label{sec:model}
The main security goal of this work is to guarantee that a Lightning party that
honestly follows the protocol never loses money. In particular, an honest
Lightning party that has sent and received a number of payments in the channel
should be able to redeem on-chain its initial channel funds plus any funds
received minus any funds sent within a bounded time after requesting channel
closure. We call this \emph{funds security}. Formally it is a property over $5$
parameters:
\begin{itemize}
\item The honest party LN protocol \texttt{HonestParty},
\item The ledger protocol $\Gamma$,
\item The party activation window $\maxwake \in \mathbb{N}$,
\item The ledger delivery window $\maxdelivery \in \mathbb{N}$.
\item The time needed for a channel to close $\channelClosingTime \in \mathbb{N}$.
\end{itemize}

We define funds security in terms of a game, played between the adversary and
the honest party. If the adversary cannot win no matter their actions, then
funds security is guaranteed. This game is defined in the
\texttt{ho\-nest\-Par\-ty\-Vs\-Ad\-ver\-sa\-ry.mlw} file.

The game state consists of the following elements: the honest party state, the
ledger state, the signature functionality state, and the collection of
timestamped, pending ledger messages. It also tracks $4$ variables: the last
time the honest party was woken ($\lastactivation \in \mathbb{N}$), whether and
when was the honest party first ordered to close its channel ($\closetime \in
\mathbb{N} \cup \{\bot\}$), the current time ($\timenow \in \mathbb{N}$), and
the expected funds of the honest party ($\expectedfunds \in \mathbb{N}$).

Since channels are two-party constructions, the other party is presumed
corrupted. We define an initial system state in which the ledger contains some
coins for each of the two parties and the channel is not yet opened.

We next describe the types of moves that the adversary can make. All actions
apart from \texttt{IncrementTime} are parametrized by additional arguments that
the adversary chooses along with the move type, as explained below.

\begin{enumerate}
  \item \texttt{SignMsg}: This message enables the adversary to sign any number 
  of messages on behalf of the corrupted party. It also models signature 
  malleability by allowing the adversary to create new signatures for messages
  already signed by the honest party (see Subsec.\ \emph{Allowing public modification of signatures}
  in~\cite{DBLP:conf/csfw/Canetti04}).
  \item \texttt{SendMsgToGamma}: The adversary generates an arbitrary message
  which is immediately processed by $\Gamma$. This models the ability of a
  \emph{rushing} adversary to add transactions to the block\-chain at any time,
  ``skiping the queue''. As we mentioned, $\Gamma$ is a parameter of the game.
  Upon receiving a message, it reads the current time, verifies any signatures
  with the signature functionality, and returns its updated state.
  \item \texttt{DeliverMsgToGamma}: This action gives the adversary control of
  the delivery of messages from \texttt{Ho\-nest\-Par\-ty} to $\Gamma$. The adversary
  chooses a message from the collection of pending messages. The message is
  removed from the collection and is immediately processed by $\Gamma$.
  \item \texttt{IncrementTime}: This action increases the current system time by
  $1$ unit. This action is only available if (i) there are no pending messages
  older than $\maxdelivery-1$ and (ii) \texttt{Ho\-nest\-Par\-ty} has been activated
  within the last $\maxwake-1$ rounds (i.e., $\timenow - \lastactivation <
  \maxwake$).
  \item \label{move:party-msg} \texttt{SendMsgToParty}: This action lets the
  adversary interact with
  \texttt{Ho\-nest\-Par\-ty} on behalf of the corrupted one. The adversary chooses a
  message to be delivered to \texttt{Ho\-nest\-Par\-ty}, which the latter handles according to its implementation. The result of the \texttt{Ho\-nest\-Par\-ty} activation is accounted for as follows. The messages that \texttt{Ho\-nest\-Par\-ty} wants to send to
  $\Gamma$ are added to the pending messages collection, \lastactivation\ is
  updated to \timenow, and \expectedfunds\ is updated if \texttt{Ho\-nest\-Par\-ty} has
  acknowledged receipt of a payment. During handling of the message,
  \texttt{Ho\-nest\-Par\-ty} may interact with the signature functionality. The
  exact format of
  messages that \texttt{Ho\-nest\-Par\-ty} expects depends on the concrete LN party
  protocol, which as mentioned before is a system parameter.
  \item \label{move:env-ctrl} \texttt{ControlEnvironment}\footnote{
    Actually, in the implementation both \texttt{SendMsgToParty} and \texttt{ControlEnvironment}
    are handled by \texttt{SendMsgToParty}, with appropriate arguments to distinguish between the two.
    We separate them here for clarity.
  }: This action lets the
  adversary interact
  with \texttt{Ho\-nest\-Par\-ty} on behalf of the system. This action is
  handled by \texttt{Ho\-nest\-Par\-ty} in exactly the same way as adversarial
  instructions (i.e.,
  \texttt{SendMsgToParty}) are. The only difference is that
  \texttt{ControlEnvironment} additionally may cause the system state
  to be directly updated. Here is an exhaustive list of all possible actions
  the system can prompt \texttt{Ho\-nest\-Par\-ty} to perform:
  \begin{enumerate}
    \item \texttt{EnvOpenChannel}: Orders \texttt{Ho\-nest\-Par\-ty} to initiate opening a
    channel with the counterparty.
    \item \texttt{TransferOnChain}: Orders \texttt{Ho\-nest\-Par\-ty} to
    transfer its funds using an on-chain transaction, i.e., by spending coins
    from a public-key account. It fulfills two functions: (i) funding of a new channel and (ii)
    direct on-chain payment to the counterparty\footnote{Our work only needs the
    former function of \texttt{TransferOnChain} --- still, the latter makes our
    model more complete for essentially no added code complexity.}. The system
    then decreases the party's \expectedfunds\ accordingly.
    \item \texttt{TransferOnChannel}: Similar to above, but using an already open
    channel. Once again, the party's \expectedfunds\ is decreased.
    \item \texttt{CloseNow}: Orders \texttt{Ho\-nest\-Par\-ty} to initiate the closing
    procedure of a previously opened channel. The variable \closetime\ is set to
    \timenow.
    \item \texttt{JustCheckGamma}: Wakes up \texttt{Ho\-nest\-Par\-ty} to give it the
    chance to do any recurrent bookkeeping. This is needed to satisfy
    the periodic \texttt{Ho\-nest\-Par\-ty} wake-up requirement, even when there is no concrete
    instruction for the party.
  \end{enumerate}
\end{enumerate}

Those actions ensure that the interactions between the adversary and the honest party are modeled accurately.
In particular, \expectedfunds\ is tracked by the experiment (and not simply reported by the honest party),
which guarantees that the value of \expectedfunds\ matches its intuitive meaning:
it decreases whenever the party is asked to pay the counterparty
and increases every time the party acknowledges receiving a payment.

Last but not least, let us define the adversary's winning condition. It is also
formally defined as \texttt{adversaryWinningState}
in \texttt{ho\-nest\-Par\-ty\-Vs\-Ad\-ver\-sa\-ry.mlw}.

\begin{definition}[Funds Security]
\label{def:funds-sec}
The winning state for an adversary is one that satisfies the following:
\begin{enumerate}
  \item $\closetime \neq \bot$,
  \item $\timenow \geq \closetime + \channelClosingTime$,
  \item The honest party has less coins than \expectedfunds\ on-chain (i.e., not
  counting those in the channel), as output by $\Gamma$ when its function
  \texttt{im\-me\-di\-ate\-A\-mount\-On\-Cha\-in} is called.
\end{enumerate}

We say that a system parametrized with the honest party LN protocol
\texttt{Ho\-nest\-Par\-ty}, the ledger
protocol $\Gamma$, \maxwake, \texttt{deltaNet}, and
\texttt{channelClosingTime} achieves \emph{funds security} if no adversary can reach the
winning state, no matter which sequence of actions it follows.
\end{definition}

Realistic LN implementations use practical constructions of digital signatures,
therefore they can only achieve Funds Security against \emph{probabilistic
po\-ly\-no\-mi\-al-ti\-me} (PPT) adversaries. As discussed earlier however, we model
digital signatures as an ideal functionality, the security of which cannot be
broken even by unbounded or non-computable adversaries. In the context of Why3
we model the adversary as unbounded for simplicity, but our ultimate security
guarantee is with respect to practical digital signature constructions and
therefore against bounded, PPT adversaries. Weakening the adversary from
unbounded to PPT strictly shrinks the set of admissible adversaries, therefore
the security guarantee obtained by the Why3 model is directly transferable to
the bounded adversarial setting.

Let \texttt{channelTimelock} be the timelock after which a party can reclaim its funds
from a commitment transaction. \texttt{channelTimelock} is a parameter of the
\texttt{HonestParty} protocol.
For any $\maxwake$, \texttt{deltaNet}, \texttt{channelClosingTime} $\in$
$\mathbb{N}:$ \texttt{channelClosingTime} $\geq$ $3$ $\cdot$
\texttt{deltaNet} $+$ $2$ $\cdot$ \texttt{deltaWake} $+$
\texttt{chan\-nel\-Ti\-me\-lock} $+$ $1$
and $\texttt{channelTimelock} \geq \maxwake + \texttt{deltaNet} + 1$,
we provide a concrete
implementation of $\Gamma$ that models a subset of the functionality of Bitcoin (Sec.~\ref{sec:bitcoin}), as well as an implementation of a
subset of LN (Subsec.~\ref{subsec:flow}) such that funds security is guaranteed, as we formally verify (the main security result is 
lemma \texttt{honestPartyWins} in \texttt{ho\-nest\-Par\-ty\-Vs\-Ad\-ver\-sa\-ry.mlw}).

\section{Modeling Bitcoin}
\label{sec:bitcoin}

We use a high-level modeling of the ledger called $\Gamma$ which captures the
fragment of Bitcoin that is relevant to Lightning channels. In this section, we
give a brief introduction to Bitcoin operation and then explain how we simplify
it in order to express just what is necessary for the protocol.

In practical implementations, Bitcoin transactions are organized in a
\emph{directed acyclic graph}
(DAG). Each transaction is a node with at least one \emph{input} and one
\emph{output}. Each input is connected to exactly one output and each output to
at most one input. Each output specifies the number of coins it contains. At any
point in time, the DAG has some outputs that are not connected to any input:
these are the \emph{unspent transaction outputs} (UTXOs) and model all available
coins. A new transaction can be added to the DAG if all its inputs are connected
to UTXOs and the sum of the coins of its outputs are equal to the sum of
the coins of the outputs that its inputs spend --- this guarantees new coins are
not created out of thin air. Furthermore, outputs contain \emph{Script} that
specify spending semantics. Script is a simple non-Turing-complete language with
a limited expressiveness, including signature verification, hash checking and
time checking. Each output locks its coins by specifying a Script statement. It
can later be unlocked by providing an input with a corresponding witness. When a
valid transaction is added to the DAG, the outputs it spends are not UTXOs
anymore and its own outputs become UTXOs.

Our modeling keeps track of the set of transaction outputs, as well as whether
they are unspent. It forgoes the DAG and the linearization of the transaction
history. We limit our attention to a fixed set of scripts:
\begin{itemize}
  \item The public key account, known as ``pay-to-public-key-hash'' (P2PKH) in
  Bitcoin implementations\footnote{Similarly to~\cite{DBLP:conf/csfw/Canetti04},
  our modeling foregoes the use of public keys, opting for destination
  identifiers instead --- this simplifies the model by avoiding the explicit
  formalization of the mechanics of cryptographic signatures.},
  \item The 2-out-of-2 multisig --- this is used to store the coins of an open
  Lightning channel during normal operation,
  \item The commitment transaction conditional output --- this is used to unilaterally close an open
  channel.
\end{itemize}
% In our implementation the names of the three different types of scripts are
% \texttt{PublicKeyAccount}, \texttt{Normal}, and \texttt{DisputeOpen}.

Bitcoin has no official specification --- this role is instead fulfilled by the
reference implementation. A formal model for Bitcoin is presented
in~\cite{DBLP:conf/fc/AtzeiBLZ18}. In the current work we intentionally avoid
modeling parts of Bitcoin execution that are irrelevant to the Lightning
participants. More specifically, we ignore all public keys other than those of
the two channel parties and any script other than the aforementioned ones. We
are confident that our simplified modeling is applicable to a full model of all
Bitcoin transitions~\cite{DBLP:conf/fc/AtzeiBLZ18} when non-Lightning
transactions are filtered out. We however leave proof of that our model is an
abstract interpretation of~\cite{DBLP:conf/fc/AtzeiBLZ18} as future work.

The aforementioned scripts are enough to model all possible Lightning
interactions. In particular, each script is used for the following transitions:
\begin{itemize}
  \item P2PKH allows simple transfers of coins by providing the owner's
  signature. There are three valid destinations of such transfers: the
  counterparty (for direct payments), a 2-out-of-2 multisig (for funding the
  channel), and a special destination for burning coins. The latter is useful
  for the completeness of our formal model but is not used in the actual
  Lightning protocol.
  \item The 2-out-of-2 multisig can be spent in exactly one way, namely by a
  commitment transaction that has been signed by both counterparties. Each
  commitment transaction has two outputs: the aforementioned commitment transaction conditional output
  and a P2PKH of the publisher's counterparty.
  \item The commitment transaction conditional output can be spent in two ways: by its publisher after a
  fixed timeout, or at any time by its counterparty if it presents a valid
  revocation transaction.
  \item The adversarial counterparty can create arbitrary P2PKH accounts. The
  adversary is allowed to do this without explicitly spending any output.
\end{itemize}
We refer the reader to our implementation for further details (\texttt{gamma.mlw}).

For simplicity, we do not model on-chain fees in our work (this is in line with~\cite{DBLP:conf/fc/AtzeiBLZ18}).
However, we believe that extending our work to include fees is conceptually straightforward.
See Section~\ref{sec:conclusion} for more details.

Note that in the
full Bitcoin a 2-out-of-2 multisig can be spent in an arbitrary way as long
as both implicated parties agree. Since our security experiment assumes that at
least one of the two channel counterparties is honest\footnote{Assuming that
both parties are malicious and no party is honest is arguably not interesting,
as security guarantees are only meaningful for honest parties.}, only the
aforementioned way to spend the 2-out-of-2 multisig is possible in our
protocol. We deduce that this constraint of 2-out-of-2 multisigs does not
artificially restrict the adversarial capabilities.
\section{Proof Strategy}
\label{sec:proof}

LN consists of multiple moving parts that are interconnected in intricate ways.
In order to manage the complexity of these interactions, we adhere to the
following principle. The system does not need to maintain the
entire history of the past states, neither for the honest LN party nor for
$\Gamma$. All useful knowledge on past states is instead compressed into
constant-size invariants. The alternative would be to keep track of the entire
execution history, which would be more aligned with human reasoning about
distributed systems. Unfortunately, quantification over time complicates
formally verified proofs, as it would require a number of interconnected
inductive proofs. Our approach is more natural in the context of Hoare logic, as
we adhere to a single induction step per protocol step.

\subsection{Digital Signatures Ideal Functionality}
\label{subsec:signatures}

From a cryptographic perspective, the security of any digital signatures
construction is only \emph{computational}, i.e., it can be broken by a
sufficiently powerful adversary. As a standard, cryptographic literature only considers
PPT adversaries. This is normally necessary because of the underlying
computational assumptions, which are breakable by an unbounded adversary. In
fact, replacing the aforementioned ideal functionality with a real signature
scheme is only secure against such PPT adversaries. Unfortunately, quantifying
over all PPT machines has historically proven exceedingly
difficult~\cite{DBLP:conf/pldi/LiaoHM19} in the context of formal verification.
We would thus like to categorically
rule out the exponentially small probability of a successful forgery in a
principled manner. We achieve this in two steps: Firstly, we allow the adversary
in Why3 to be any function (even non-computable) using the \texttt{any} keyword of Why3.
We then replace the signatures construction with an
\emph{idealized} digital signatures functionality (\texttt{signaturesFunctionality.mlw})
which never permits forgeries
(not even with \emph{negligible} probability), as defined
in~\cite{DBLP:conf/csfw/Canetti04}. At a high level, this
functionality plays the role of a ``trusted signatures server'' that is
common for all parties. Parties request the signing and verification of
messages. The functionality stores the message-party
pair on signature request and responds whether the given message-party
pair has been stored on verification request. Crucially, the
functionality is incorruptible, making forgeries impossible even by a
non-computable adversary.

A salient question is: why is the replacement of the construction with the
functionality valid? The answer, coming from the \emph{simulation-based
cryptographic paradigm}, employs an \emph{indistinguishability
argument}, according to which \emph{every} external PPT observer is unable to distinguish between interactions with
the construction versus the functionality, except with negligible probability.
Indeed, our ultimate security guarantee is provided with respect to a realistic
signatures scheme (not an ideal functionality) and a PPT adversary (not an
unbounded one). Using a non-computable adversary is merely a proof technique,
necessary to tractably model the adversarial behavior in Why3. The
indistinguishability argument enables us to transfer the ideal-world security
property (formally verified by Why3) to the real world of practical digital
signatures and PPT adversaries.

\subsection{Encapsulation of Implementation-aware Properties}
\label{subsec:party-interface}
Our work includes a novel implementation of a subset of the LN mechanism, an
implementation of Bitcoin on which our LN implementation relies,  as well as a
formally verified proof of the \emph{funds security} of our implementation. We
chose to organize our proof so as to clearly separate the
particularities of our custom LN implementation on the one hand and the general
logic that every secure LN implementation must uphold on the other. Our proof is
therefore separated into three parts: an implementation-specific part
(\texttt{honestPartyType.mlw} and \texttt{honestPartyImplementation.mlw})
which is aware of the complexities and details of our LN implementation, an
abstract specification which encapsulates a set of simpler invariants that any
secure LN implementation must uphold (\texttt{partyInterface.mlw}), and an
implementation-agnostic part of the proof that depends only on the specification
(\texttt{honestPartyVsAdversary.mlw}).

Our abstract specification focuses on the high-level properties that clients
have to satisfy in order to uphold security, such as the need to publish
on-chain a known revocation transaction as soon as it becomes valid. This is
very different from BOLT, which specifies the required behavior of clients,
(e.g., reactions to specific messages), but does not stipulate logical
properties. For more details on the interface, see Appx.~\ref{app:party-interface} or 
\texttt{partyInterface.mlw}.

\subsection{\texttt{goodSimpleParty}}
\label{subsec:partyview}

As our \texttt{HonestParty} LN protocol
implementation progressed, it became increasingly complex and highly coupled
with the rest of the system. Changes to low-level details of
\texttt{HonestParty} LN
protocol would ripple to the invariants and from there to the rest of the
system. To mitigate this issue, we decided to introduce a simple, implementation-agnostic 
invariant that captures the essence of the properties that any secure LN
implementation should satisfy (\texttt{goodSimpleParty} in \texttt{partyInterface.mlw}).
This invariant is aware of only a subset of the information held by the 
\texttt{HonestParty} implementation, such as the received revocations and the commitment
transaction that can be used to close the channel. This is the data that any
reasonable LN implementation should be able to produce. (It is represented 
by the type \texttt{simplePartyT} in \texttt{partyInterface.mlw}.)

Unfortunately, the invariant \texttt{goodSimpleParty} cannot be proven directly by induction --- one cannot prove that the invariant is preserved by the \texttt{HonestParty} transitions without additional dependencies. For this reason we also define a stronger,
implementation-aware invariant 
(\texttt{partyInvariant} in \texttt{ho\-nest\-Par\-ty\-Ty\-pe.mlw}) that implies \texttt{goodSimpleParty}, and that could be proven directly by induction.
This invariant is not visible to the rest of the system and is only used internally
in the implementation of the \texttt{HonestParty}. To highlight the difference in complexity
let us mention that the core logic of \texttt{partyInvariant} spans
approximately $65$ lines of code, whereas that of \texttt{good\-Sim\-ple\-Pa\-rty} is
just around $15$ lines.

\subsection{Funds Ownership}
\label{subsec:ownership}
In this subsection we discuss what it means for an \texttt{HonestParty}
to own coins in $\Gamma$. This is a subtle point, since, from the perspective of
the chain, all channel funds are locked behind a 2-of-2 multisig, which usually
does not imply exclusive ownership. However, an
honest channel party should be able to consider some of these funds as its own.
As a further example of funds ownership complications, consider an on-chain output
controlled by a single public key \textit{pk} and a signed, valid, published
transaction that spends this output but is not (yet) part of the chain. These
funds should not be considered as owned by the holder of the private key of
\textit{pk}. Our approach to resolving such ambiguities to ownership is to consider
as owned those coins that a party can reliably and unilaterally move to its
public key account --- this way the tension in both of these
scenarios vanishes.

To formalize this concept of ownership, consider a specific state of the
Bitcoin ledger. Let an \emph{Extractor} be a party which has some local state (e.g.,
private keys, signatures received by others) and can communicate with the
ledger. We also consider an \emph{Obstructor} with access to the ledger and the
ability to take any action of the adversary, as described in
Sec.~\ref{sec:model}. The Extractor has a single ID with respect to the
signature functionality, whereas the Obstructor controls all other IDs. The
Extractor's goal is to maximize the funds it can extract from the ledger into
its private account(s),
whereas the Obstructor tries to minimize them. Let \emph{extractable value}
be the value of such a game when both parties play optimally.
%--- this value is well-defined, as it is the minimax value of a zero-sum
%game~\cite{gametheory}.
We can now define the value owned by a given party
at any instance of the execution of some Bitcoin-based protocol by identifying
this party with the Extractor (thus allowing the party to diverge from the
protocol) and the rest of the parties (along with the adversary) with the
Obstructor. We then say that the funds owned by the party are equal to the extractable
value. Observe that this definition of ownership does not depend on any specific
protocol thus it can be used in settings other than LN.

Next, we discuss how the concept of ownership is leveraged in our proof of funds
security. We say that the \texttt{HonestParty} has \emph{solvency} at some point
during the execution of the funds security game if the funds it owns are at
least equal to \expectedfunds\ (Sec.~\ref{sec:model}). Solvency has the intuitively appealing property that it
refers to any given moment \emph{during} the execution, as opposed to funds
security which only concerns the \emph{final state}. It is not hard to see that solvency
implies funds security when the channel is closed. More precisely, funds
security follows from solvency and the fact that the \texttt{HonestParty} can
close the channel in bounded time. The proof of the latter fact is relatively 
straightforward (it is formalized as lemma \texttt{closingWorks} in \texttt{honestPartyVsAdversary.mlw}).
The solvency part of the proof is more involved and it is discussed in Subsec.~\ref{subsec:evaluator-consistency}.

Finally, let us point out an important technical detail:
Although the game-based definition of ownership is, in our opinion, both intuitive and theoretically appealing,
it is also challenging to compute and reason about, as it involves evaluating an arbitrary min-max tree.
Therefore, in our proof, we replace funds ownership with a more straightforward (i.e., free of min-max trees) but less intuitive \texttt{Evaluator} function
(implemented as \texttt{partyExpectationsFull} in \texttt{gamma.mlw}),
which provides a pessimistic under-approximation of the funds owned by the
party. For further details, see Appx.~\ref{sec:evaluator}.
Similarly, we replace solvency with \emph{evaluator solvency}, which states that the \texttt{Evaluator}
never drops below \expectedfunds\ during the execution of the protocol.

\subsection{Evaluator Solvency Overview}
\label{subsec:evaluator-consistency}

As discussed in Subsec.~\ref{subsec:ownership}, a central point of our effort is
proving the preservation of evaluator solvency across state transitions.
In this section we focus on the crux of the proof, which lies in 
the transitions of $\Gamma$ and the \texttt{HonestParty}.

For the transitions of $\Gamma$, it is sufficient to
prove that the \texttt{Evaluator} does not decrease (as \expectedfunds{}
does not change during such transitions). This is not hard to prove
as the \texttt{Evaluator} mirrors the crucial aspects of funds
ownership which, by definition, does not decrease during transitions of $\Gamma$.
(The proof is formalized as \texttt{gammaProcessFreshMsgPreservesEvaluatorG} in \texttt{gamma.mlw}.)

The transitions of the \texttt{Ho\-nest\-Par\-ty} are harder to handle for two reasons:
Firstly, they may change both the \texttt{Evaluator} and the \expectedfunds{} (see moves~\ref{move:party-msg} and~\ref{move:env-ctrl} in
Sec.~\ref{sec:model}). Secondly, the transitions of the \texttt{Ho\-nest\-Par\-ty} are more involved than the ones of $\Gamma$.
To help us with the proof, we define a relation \texttt{goodTransition} (in \texttt{par\-ty\-In\-ter\-face\-.mlw}), which connects the states of
the \texttt{simplePartyT} before and after the transition, as well as the messages sent by the party to $\Gamma$ during the transition.
This relation abstracts the properties that any LN implementation must guarantee to ensure evaluator solvency.
It concerns the obligations of \texttt{HonestParty} during a state transition. For example, in
case of dispute, the party has to publish a suitable stored revocation to $\Gamma$.

The relation \texttt{goodTransition} has two desirable
properties that make it useful in the proof. Firstly,
it is agnostic of the exact implementation of \texttt{HonestParty},
which allows us to separate the proof from the implementation details.
This is achieved by splitting the proof into two steps:
in the first step we prove that any party update that satisfies \texttt{goodTransition} preserves
evaluator solvency (formalized as \texttt{to\-tal\-E\-va\-lu\-a\-tor\-Mi\-nus\-To\-tal\-Ba\-lance\-Mo\-no\-tone}
in \texttt{par\-ty\-In\-ter\-face\-.mlw}), and in the second step we show that
the transitions of \texttt{HonestParty} satisfy \texttt{good\-Tran\-si\-tion}
(formalized as a postcondition of \texttt{partyProcessMsg} in \texttt{ho\-nest\-Par\-ty\-In\-ter\-ac\-tions\-.mlw}).
The other desirable property of \texttt{good\-Tran\-si\-tion} is its transitivity,
which allows us to split the second step of the proof into smaller,
modular substeps.
\section{Conclusion \& Future Work}
\label{sec:conclusion}

In this work we successfully modeled Lightning~\cite{lightning} channels using
first-order logic in Why3 and proved funds security for honest channel parties.
First-order logic allowed us to express naturally the security property
and functionality of Lightning channels, as well as the low-level invariants
encountered during formalization. During the proof effort, we realized that an in-depth understanding of funds ownership was necessary,
which led us to define it robustly and formally. We believe that this approach
can be reused in similar efforts in the future. Another relevant outcome is the
\texttt{partyInterface} design, which was the result of a number of
modularization attempts. It cleanly separates LN party implementation details
with the interface it should provide, simplifying the proof.

There are a number of future directions for strengthening our  modelling and
architecture. To begin with, we believe that our work could be extended to
two-party channels supporting HTLCs with a reasonable amount of effort. This
will be a significant stepping stone towards formalizing the security of
real-world LN implementations. Moreover, our modeling of Bitcoin can be improved
to bring it to parity with already existing formalizations such
as~\cite{DBLP:conf/fc/AtzeiBLZ18}. We expect that this task also needs a
reasonable effort, but could face more
unexpected roadblocks than adding HTLCs.

Another way of improving our Bitcoin model is to include transaction fees,
which, for simplicity, we currently do not model. This would require the
honest parties to set aside some on-chain funds to pay any fees that may arise.
These funds would not be transferable by the \texttt{TransferOnChain} environment order.
Thanks to the small maximum number of per-channel on-chain transactions (at most $3$)
and given a (weak) assumption of a maximum per-transaction fee, the amount to set aside is fixed.
Thus, funds security could also be proved in the presence of transaction fees.

Last but not least, a more ambitious
project would be to extend the security analysis to multi-hop payments over a
network of LN channels. To that end, new ideas and insights would be required.

% TODO: uncomment
\subsubsection*{Acknowledgements.}
This work has been partly supported by the
European Research Council (ERC) under the European Union's Horizon 2020 innovation program
(grant PROCONTRA-885666). Furthermore, this work was partly supported by the German Research Foundation (DFG) via the DFG CRC 1119 CROSSING (project S7), by the German Federal Ministry of Education and Research and the Hessen State Ministry for Higher Education, Research and the Arts within their joint support of the National Research Center for Applied Cybersecurity ATHENE.

\bibliography{src/references}
\appendix
%\import{./src/}{signatures}
\section{System and proof components}
\label{sec:components}
In the following picture, we present the key components of our system and proofs.
The solid boxes represent parts of the implementation --- they are covered in Subsec.~\ref{subsec:architecture-overview}.
The two cloud-shaped boxes represent the two main parts of the proof: \texttt{EvaluatorSolvencyPreservation} (Subsec.~\ref{subsec:evaluator-consistency})
and \texttt{FundsSecurity} (Sec.~\ref{sec:model}).
\custompicc{pictures/diagram}{0.8}

\section{\texttt{Evaluator}}
\label{sec:evaluator}
As mentioned in Subsec.~\ref{subsec:ownership}, funds ownership is very
complicated to keep track of and even harder to reason about. In particular,
it involves calculating the minimax value that emerges from arbitrary
transaction trees with possibly conflicting transactions. We thus opted for a
much simpler, pessimistic approximation of the owned funds and implemented it in
the \texttt{Evaluator} function (namely \texttt{partyExpectationsFull} in
\texttt{gamma.mlw}). Instead of proving solvency, we prove
\emph{evaluator solvency}, i.e., that at any point during the funds security game it
holds that $\expectedfunds \geq \texttt{Evaluator}$. (The original concepts of funds
ownership and its solvency do not appear in our code.) We then show that, when the
channel is closed, evaluator solvency implies funds security
(see Lemma~\texttt{fundsSecurityAux} in \texttt{honestPartyVsAdversary.mlw} --
the assumption about the channel being closed is derived from the time assumptions
combined with Lemma~\texttt{closingWorks}).
Note that this proof strategy matches the one described in
Subsec.~\ref{subsec:ownership}, with funds ownership replaced
by \texttt{Evaluator}.

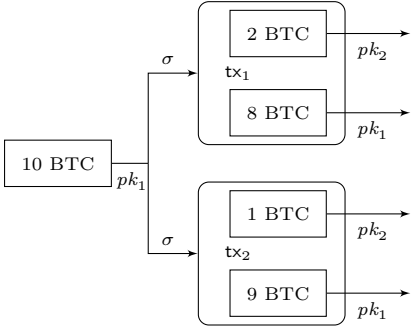
\begin{figure}[htbp!]
\begin{center}
\begin{tikzpicture}[auto, node distance=1.2 cm] % TODO to shorten: tweak `node distance`
  \node[font=\scriptsize] (initialcoins) {$10$ BTC};
  \node[output, fit=(initialcoins)] (initialcoinsbox) {};
  \coordinate[right of=initialcoinsbox] (inputmid);

  \draw (initialcoinsbox) -- (inputmid) node[labelBelow_small, font=\scriptsize, pos=0.55pt] {$\pk{1}$};

  \node[right of=inputmid, above of=inputmid, font=\scriptsize] (tx1label) {$\tx{1}$};
  \node[node distance=15pt, right of=tx1label, above of=tx1label, font=\scriptsize] (tx1payout) {$2$ BTC};
  \node[output, fit=(tx1payout)] (tx1payoutbox) {};
  \coordinate[node distance=50pt, right of=tx1payoutbox] (tx1payoutrecipient);
  \path [arrow] (tx1payoutbox) -- (tx1payoutrecipient) node[labelBelow_small, font=\scriptsize, pos=0.55pt] {$\pk{2}$};
  \node[node distance=15pt, right of=tx1label, below of=tx1label, font=\scriptsize] (tx1change) {$8$ BTC};
  \node[output, fit=(tx1change)] (tx1changebox) {};
  \coordinate[node distance=50pt, right of=tx1changebox] (tx1changerecipient);
  \path [arrow] (tx1changebox) -- (tx1changerecipient) node[labelBelow_small, font=\scriptsize, pos=0.55pt] {$\pk{1}$};
  \node[TX, fit=(tx1label)(tx1payoutbox)(tx1changebox)] (tx1box) {};

  \path [arrow] (inputmid) |- (tx1box) node[labelAbove_small, font=\scriptsize, pos=0.7pt] {$\sigma$};

  \node[right of=inputmid, below of=inputmid, font=\scriptsize] (tx2label) {$\tx{2}$};
  \node[node distance=15pt, right of=tx2label, above of=tx2label, font=\scriptsize] (tx2payout) {$1$ BTC};
  \node[output, fit=(tx2payout)] (tx2payoutbox) {};
  \coordinate[node distance=50pt, right of=tx2payoutbox] (tx2payoutrecipient);
  \path [arrow] (tx2payoutbox) -- (tx2payoutrecipient) node[labelBelow_small, font=\scriptsize, pos=0.55pt] {$\pk{2}$};
  \node[node distance=15pt, right of=tx2label, below of=tx2label, font=\scriptsize] (tx2change) {$9$ BTC};
  \node[output, fit=(tx2change)] (tx2changebox) {};
  \coordinate[node distance=50pt, right of=tx2changebox] (tx2changerecipient);
  \path [arrow] (tx2changebox) -- (tx2changerecipient) node[labelBelow_small, font=\scriptsize, pos=0.55pt] {$\pk{1}$};
  \node[TX, fit=(tx2label)(tx2payoutbox)(tx2changebox)] (tx2box) {};

  \path [arrow] (inputmid) |- (tx2box) node[labelAbove_small, font=\scriptsize, pos=0.7pt] {$\sigma$};
\end{tikzpicture}
\end{center}
\caption{Conflicting transactions. $\tx{1}$ pays $2$ coins from
$\pk{1}$ to $\pk{2}$, whereas $\tx{2}$ pays $1$ coin from $\pk{1}$ to
$\pk{2}$. Since they spend the same output, at most one can be
included on-chain. Funds ownership takes into account that the two
transactions are mutually exclusive and only considers the maximum
payout, $2$, concluding that $\pk{1}$ owns $8$ coins. In contrast,
the \texttt{Evaluator} considers the sum of the two payouts,
$1+2=3$, concluding that $\pk{1}$ owns $7$ coins.}
\label{fig:mutually-exclusive}
\end{figure}

Let us now elaborate on the difference between funds ownership and
\texttt{E\-va\-lu\-a\-tor}. Interestingly, their main difference lies in the way they
treat the simplest Bitcoin outputs, i.e., those locked with a single public key.
We will introduce the difference by example. Consider a simple, on-chain output
with $10$ coins and two competing, fully signed transactions that spend this
output. Neither of these transactions is part of the block\-chain. The first
transaction pays out $1$ coin and returns the other $9$ to the original public
key, whereas the second pays out $2$ coins and returns $8$
(Fig.~\ref{fig:mutually-exclusive}).
To calculate the funds owned by the original public key, we observe that, since only one of these
transactions can be added to the chain, the best strategy for the Obstructor is to
choose the transaction with the \emph{maximum} payout, which in this example is $2$ coins.
Therefore the funds owned are equal to $10-2=8$ coins.
In contrast, the \texttt{Evaluator} does not take into
account the fact that the two transactions are mutually exclusive. It instead
calculates the total payout as the \emph{sum} of the two payouts, i.e., $1+2=3$
coins. Therefore the \texttt{Evaluator} outputs $10-1-2=7$ coins\footnote{The keen
reader might here observe that the \texttt{Evaluator} may, in the general case,
output negative values. We clarify that, even though this runs contrary to
intuition, it does not constitute a problem.}. This difference does not affect
the \texttt{HonestParty} protocol, as the latter never signs two conflicting
transactions that spend its public key output.

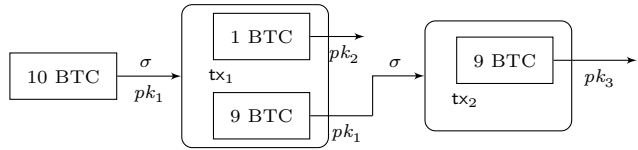
\begin{figure}[htbp!]
\begin{center}
\begin{tikzpicture}[auto, node distance=1.2 cm]
  \node[font=\scriptsize] (initialcoins) {$10$ BTC};
  \node[output, fit=(initialcoins)] (initialcoinsbox) {};

  \node[node distance=60pt, right of=initialcoinsbox, font=\scriptsize] (tx1label) {$\tx{1}$};
  \node[node distance=15pt, right of=tx1label, above of=tx1label, font=\scriptsize] (tx1payout) {$1$ BTC};
  \node[output, fit=(tx1payout)] (tx1payoutbox) {};
  \coordinate[node distance=39pt, right of=tx1payoutbox] (tx1payoutrecipient);
  \path [arrow] (tx1payoutbox) -- (tx1payoutrecipient) node[labelBelow_small, font=\scriptsize, pos=0.62pt] {$\pk{2}$};
  \node[node distance=15pt, right of=tx1label, below of=tx1label, font=\scriptsize] (tx1change) {$9$ BTC};
  \node[output, fit=(tx1change)] (tx1changebox) {};
  \coordinate[node distance=42pt, right of=tx1changebox] (tx1changerecipient);
  \draw (tx1changebox) -- (tx1changerecipient) node[labelBelow_small, font=\scriptsize, pos=0.59pt] {$\pk{1}$};
  \node[TX, fit=(tx1label)(tx1payoutbox)(tx1changebox)] (tx1box) {};

  \draw [arrow] (initialcoinsbox) -- (tx1box)
    node[labelBelow_small, font=\scriptsize, pos=0.5pt] {$\pk{1}$}
    node[labelAbove_small, font=\scriptsize, pos=0.45pt] {$\sigma$}
  ;

  \coordinate[node distance=9pt, below of=tx1label] (tx2helper);
  \node[node distance=92pt, right of=tx2helper, font=\scriptsize] (tx2label) {$\tx{2}$};
  \node[node distance=15pt, right of=tx2label, above of=tx2label, font=\scriptsize] (tx2payout) {$9$ BTC};
  \node[output, fit=(tx2payout)] (tx2payoutbox) {};
  \coordinate[node distance=50pt, right of=tx2payoutbox] (tx2payoutrecipient);
  \path [arrow] (tx2payoutbox) -- (tx2payoutrecipient) node[labelBelow_small, font=\scriptsize, pos=0.55pt] {$\pk{3}$};
  \node[TX, fit=(tx2label)(tx2payoutbox)] (tx2box) {};

  \draw [arrow] (tx1changerecipient) |- (tx2box) node[labelAbove_small, font=\scriptsize, pos=0.72pt] {$\sigma$};
\end{tikzpicture}
\end{center}
\caption{$\tx{2}$ is reachable, since $\tx{1}$ is signed. Both funds ownership
and the \texttt{Evaluator} agree that $\pk{1}$ has $0$ coins.}
\label{fig:reachable}
\end{figure}

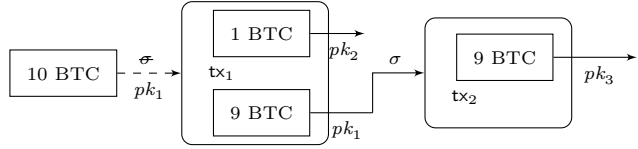
\begin{figure}[htbp!]
\begin{center}
\begin{tikzpicture}[auto, node distance=1.2 cm]
  \node[font=\scriptsize] (initialcoins) {$10$ BTC};
  \node[output, fit=(initialcoins)] (initialcoinsbox) {};

  \node[node distance=60pt, right of=initialcoinsbox, font=\scriptsize] (tx1label) {$\tx{1}$};
  \node[node distance=15pt, right of=tx1label, above of=tx1label, font=\scriptsize] (tx1payout) {$1$ BTC};
  \node[output, fit=(tx1payout)] (tx1payoutbox) {};
  \coordinate[node distance=39pt, right of=tx1payoutbox] (tx1payoutrecipient);
  \path [arrow] (tx1payoutbox) -- (tx1payoutrecipient) node[labelBelow_small, font=\scriptsize, pos=0.62pt] {$\pk{2}$};
  \node[node distance=15pt, right of=tx1label, below of=tx1label, font=\scriptsize] (tx1change) {$9$ BTC};
  \node[output, fit=(tx1change)] (tx1changebox) {};
  \coordinate[node distance=42pt, right of=tx1changebox] (tx1changerecipient);
  \draw (tx1changebox) -- (tx1changerecipient) node[labelBelow_small, font=\scriptsize, pos=0.59pt] {$\pk{1}$};
  \node[TX, fit=(tx1label)(tx1payoutbox)(tx1changebox)] (tx1box) {};

  \draw [arrow, dashed] (initialcoinsbox) -- (tx1box)
    node[labelBelow_small, font=\scriptsize, pos=0.5pt] {$\pk{1}$}
    node[labelAbove_small, font=\scriptsize, pos=0.45pt] {\sout{$\sigma$}}
  ;

  \coordinate[node distance=9pt, below of=tx1label] (tx2helper);
  \node[node distance=92pt, right of=tx2helper, font=\scriptsize] (tx2label) {$\tx{2}$};
  \node[node distance=15pt, right of=tx2label, above of=tx2label, font=\scriptsize] (tx2payout) {$9$ BTC};
  \node[output, fit=(tx2payout)] (tx2payoutbox) {};
  \coordinate[node distance=50pt, right of=tx2payoutbox] (tx2payoutrecipient);
  \path [arrow] (tx2payoutbox) -- (tx2payoutrecipient) node[labelBelow_small, font=\scriptsize, pos=0.55pt] {$\pk{3}$};
  \node[TX, fit=(tx2label)(tx2payoutbox)] (tx2box) {};

  \draw [arrow] (tx1changerecipient) |- (tx2box) node[labelAbove_small, font=\scriptsize, pos=0.72pt] {$\sigma$};
\end{tikzpicture}
\end{center}
\caption{$\tx{2}$ is unreachable, since $\tx{1}$ is unsigned. Both definitions
do not take the payout of the invalid $\tx{1}$ into account. Nevertheless,
according to funds ownership, $\pk{1}$ has $10$ coins, but the
\texttt{Evaluator} ignores the unreachability of $\tx{2}$ and thus decides that
$\pk{1}$ has $10-9=1$ coin.}
\label{fig:unreachable}
\end{figure}

As another example, consider an on-chain output $O_1$ locked by a single public key
$\pk{}$ with $10$ coins and two valid, signed transactions that are not yet
on-chain. The first, $\tx{1}$, pays out $1$ coin from $O_1$ and returns the
remaining $9$ coins to $O_2$, also controlled by $\pk{}$. The second, $\tx{2}$,
pays out the remaining $9$ coins from $O_2$ (Fig.~\ref{fig:reachable}). In this
case, funds ownership and
\texttt{Evaluator} agree that $\pk{}$ has $0$ coins. This example
highlights how both funds ownership and the \texttt{Evaluator} take into account
transactions that spend outputs that are not yet on-chain. Now consider the same
scenario but where $\tx{1}$ is instead \emph{not} signed
(Fig.~\ref{fig:unreachable}). In this case, both funds
ownership and the \texttt{Evaluator} ignore $\tx{1}$. However, funds ownership
takes into account the fact that $O_2$ is \emph{unreachable}, i.e., it
cannot appear on-chain without the missing signature, and decides that $10$ coins
are owned by $\pk{}$. On the other hand, the \texttt{Evaluator} does \emph{not}
take reachability into account. It considers $\tx{2}$ as potentially valid and thus outputs $10-9=1$. Once again,
the \texttt{HonestParty} protocol is not affected by this discrepancy, since it
never signs transactions that spend unreachable outputs that are locked by a
single public key.

As these examples imply, funds ownership is calculated very differently to the
\texttt{Evaluator} output. In particular, funds ownership is calculated by
following all possible paths of signed transactions, whether or not they are
on-chain, and choosing the one that corresponds to the minimum coins for the
public key in question. On the other hand, the \texttt{Evaluator} first
calculates the total on-chain coins that belong to the public key and then
subtracts the sum of all funds that are paid out to other public keys by pending
transactions which are signed by this key.
This causes mutually exclusive and invalid transaction paths to be also counted
as payouts, but greatly simplifies the proof effort by not considering
transaction paths. We note that \texttt{HonestParty} and other useful protocols
over Bitcoin never sign conflicting transaction paths, therefore in practice
funds ownership and the \texttt{Evaluator} output coincide. Importantly, the
\texttt{Evaluator}-based approach is of independent interest, as it can be used
to analyze other Bitcoin protocols, as long as the above considerations are
taken into account.

Let us now discuss how the \texttt{Evaluator} handles LN-specific outputs and
transactions. Contrary to the case of simple on-chain coin transfers, LN
requires parties to sign multiple mutually exclusive commitment transactions,
all of which spend the same multisig output, i.e., the output of the funding
transaction of the channel, which corresponds to the ``mutually exclusive
transactions'' example above (Fig.~\ref{fig:mutually-exclusive}). Furthermore,
each revocation transaction that an \texttt{HonestParty} signs corresponds to
the ``unreachable transaction'' example above (Fig.~\ref{fig:unreachable}).
Therefore, specifically for LN, the \texttt{Evaluator} cannot follow the same
aggressive approximation approaches, lest an honest party be usually assigned
$0$ coins (or less). Instead, during processing of LN-related transactions, the
\texttt{Evaluator} follows the funds ownership approach more closely.

First, let us discuss how the \texttt{Evaluator} treats commitment transaction outputs.
There are only two possible ways that such an output can be spent: either by a
revocation transaction, or by the public key of the commitment transaction's
publisher after a timelock. The \texttt{Evaluator} decides which of the two
paths are available to the adversary given the signatures signed by the
\texttt{HonestParty}. This decision process includes checking whether there is
still enough time so that revocations are guaranteed to be delivered before the
timelock expires\footnote{This includes reasoning both about already sent
revocations and about revocations that the \texttt{HonestParty} will certainly be able
to send in the future thanks to wake-up assumptions.}. Finally, the
\texttt{Evaluator} computes the highest value enforceable by the
\texttt{HonestParty} given the signatures that it has signed and received.

Next, we discuss how the \texttt{Evaluator} treats funding transaction outputs.
Such outputs can only be spent by a commitment transaction,
% Whenever a payment in the channel is made, each party generates a new commitment
% transaction.
so in order to calculate their value, the \texttt{Evaluator} takes into account
all possible commitment transactions available to the adversary (i.e., those
signed by the \texttt{HonestParty}) and singles out the one that minimizes the
\texttt{HonestParty}'s payout using the approach of the previous paragraph.
Since the adversary may remain idle and not publish any commitment transaction,
the \texttt{HonestParty} might need to unilaterally close the
channel. The \texttt{Evaluator} therefore returns the minimum between the
previously discussed best adversarial action and the valuation of the commitment
transaction currently stored by the \texttt{HonestParty}.

In other words, the definition of the LN-specific part of the \texttt{Evaluator}
boils down to the following observation: Since the resolution paths of LN have length at most $2$,
the \texttt{Evaluator} only needs to consider transaction trees of depth $\leq 2$.
This allows us to provide precise reasoning specifically for the case
of LN without adding too much generality and code complexity.

It is important to point out that \texttt{Evaluator} does not assume that the \texttt{Ho\-nest\-Par\-ty}
satisfies any invariants guaranteed by the \texttt{HonestParty} implementation.
For example, the \texttt{Evaluator} does not depend on the assumption that the
commitment transaction stored by the \texttt{HonestParty} is always unrevoked.
Moreover, the \texttt{Evaluator} is aware of only a subset of the information of
\texttt{HonestParty}, stored in the type \texttt{simplePartyT}
(Subsec.~\ref{subsec:partyview}): the received revocations and a commitment
transaction that can be used to close the channel. This is data that any
reasonable LN implementation should be able to produce. In particular,
the \texttt{Evaluator} does not take into account old commitment transactions
that were once stored by the \texttt{HonestParty} --- they are no longer relevant, as
the \texttt{HonestParty} has intentionally chosen to (revoke and) forget them.

In the big picture, the \texttt{Evaluator} offers a
convenient abstraction between the formal properties provided by
the low-level details of the $\Gamma$ semantics, including its state evolution over time,
and the high-level guarantees that honest LN parties should enjoy.
Without the \texttt{Evaluator}, a direct proof would require analyzing the
set of reachable states, making it dependent on the implementation of both
the \texttt{HonestParty} and $\Gamma$.
Such a proof would quickly become complicated due to the interactions
of multiple subsystems.
The \texttt{Evaluator} prevents these complications by crystallizing all $\Gamma$ state transitions
into a single, relevant property.
\section{Honest party interface}
\label{app:party-interface}
In this appendix, we describe the \emph{honest party interface} (see Subsecs.~\ref{subsec:party-interface} and ~\ref{subsec:architecture-overview} for context). 
The interface consists of a type \texttt{simplePartyT} (see Subsec.~\ref{subsec:partyview}) and two predicates:
\texttt{goodSimpleParty} (see Subsec.~\ref{subsec:partyview}) and \texttt{goodTransition} (see Subsec.~\ref{subsec:evaluator-consistency}),
which we define below. The definitions presented in this section are sometimes simplified for readability; for the full definitions, refer to \texttt{partyInterface.mlw}.
We start with the record \texttt{simplePartyT}, which contains the following fields:
\begin{enumerate}
    \item \texttt{onChainBalance} of type \texttt{amountT}. It represents the amount the party reports that it has in its 
           public key address on the blockchain.
    \item \texttt{channelBalanceExt} of type \texttt{amountT}. It represents the amount the party reports that it has in the channel.
    \item \texttt{closingChannel} of type \texttt{bool}. It is set to \texttt{true} if the party is in the process of closing the channel.
    \item \texttt{channelInfo} which is a record containing the following fields:
    \begin{enumerate}
        \item \texttt{recordOwnerG} of type \texttt{partyT}. It represents the party's identity --- either \texttt{A} or \texttt{B}.
        (We do not keep it directly in \texttt{simplePartyT} to avoid redundancy);
        \item \texttt{bestSplitReceivedG} of type \texttt{option halfSignedSplitT}. It is the commitment transaction that the party will use if it wants to close the channel, 
        equipped with the counterparty's signature. It is equal to \texttt{None} if the party is not committed to any channel.
        \item \texttt{receivedRevocationsG} of type \texttt{revokedSplitsListT}. It is a list of revoked commitment transactions that the party has received. 
        They are also equipped with the counterparty's signatures.
    \end{enumerate}
\end{enumerate}

Before discussing \texttt{goodSimpleParty}, let us introduce some terminology regarding commitment transactions. 
When a commitment transaction is submitted to blockchain, one of the parties receives a payout immediately,
while the other party can reclaim its coins after a certain timelock, if the other party does not submit a revocation.
We call those parties respectively the \emph{unconditional party} and the \emph{conditional party}. Additionally, 
we call the amount received immediately by the unconditional party the \emph{unconditional amount},
and the amount that can be reclaimed by the conditional party the \emph{conditional amount}.

The predicate \texttt{goodSimpleParty} takes a \texttt{simplePartyT} along with the state of the signature functionality as inputs, then checks that the following conditions hold:
\begin{enumerate}
    \item \label{it:balanceOur} $\mathtt{channelBalanceExt} \leq \mathtt{balanceOur}$. Here $\mathtt{balanceOur}$ is the
    conditional amount in \texttt{bestSplitReceivedG}, or $0$ if it is \texttt{None}.
    Intuitively, it is the amount that the party should be able to extract from the channel.
    \item The revocations in \texttt{receivedRevocationsG} satisfy the following conditions:
    \begin{enumerate}
        \item They are all correctly signed by the counterparty (as reported by the signature functionality);
        \item The counterparty is the conditional party in all of them.
    \end{enumerate}
    \item The \texttt{bestSplitReceivedG} satisfies the following conditions:
    \begin{enumerate}
        \item It has not been revoked (as reported by the signature functionality); 
        \item \texttt{recordOwnerG} its conditional party;
        \item It is correctly signed by the counterparty.
    \end{enumerate}
    \item Every commitment transaction $s$ signed by the \texttt{recordOwnerG} (as reported by the signature functionality) satisfies at least \emph{one} of the following conditions:
    \begin{enumerate}
        \item \texttt{receivedRevocationsG} contains a revocation for $s$;
        \item \texttt{recordOwnerG} is the unconditional party in $s$, and the unconditional amount in $s$ is at least equal to \texttt{ba\-lance\-Our} 
        (see Item~\ref{it:balanceOur});
        \item \texttt{recordOwnerG} is the conditional party in $s$, $s$ has not been revoked by \texttt{recordOwner},
              and the conditional amount in $s$ at least equal to \texttt{ba\-lance\-Our}.
    \end{enumerate}
\end{enumerate}

Next, let us discuss the predicate \texttt{goodTransition}. It is a conjunction of four sub-predicates:
\begin{enumerate}
\item \texttt{goodTransitionForChannel} --- contains the properties that guarantee evaluator consistency in the channel;
\item \texttt{goodTransitionForClosing} --- contains the properties that guarantee the timely closure of the channel (when required);
\item \texttt{goodTransitionForChain} --- contains the properties that guarantee evaluator consistency on-chain;
\item \texttt{goodTransitionAccounting} --- contains a few general properties that guarantee the consistency of the party's internal bookkeeping.
\end{enumerate} 

In the interest of brevity, we only describe \texttt{goodTransitionForChannel} --- for the rest of the predicates, please refer to \texttt{partyInterface.mlw}.
First, let us introduce some more terminology:
Observe that a channel is uniquely identified by any of its commitment transactions.
We say that a party \emph{tracks} a channel if its \texttt{best\-Split\-Re\-ceived\-G} contains 
a commitment transaction for that channel.\\

The predicate \texttt{goodTransitionForChannel} takes as its primary inputs a \texttt{sim\-ple\-Par\-tyT} before and after
a transition. Additionally, it takes as inputs the system state prior to the transition (including the blockchain state and the signature functionality)
and a record describing how the party influences the system state (including messages sent to the blockchain and updates to the signature functionality).
It then checks that the following conditions hold:
\begin{enumerate}
    \item The \texttt{recordOwnerG} did not change during the transition.
    \item If the party was tracking a channel before the transition, then it is still tracking the same channel after the transition.
    \item The party did not forget any revocation from \texttt{receivedRevocationsG}.
    \item If the party traced a channel in the \texttt{DisputeOpen} state, it did not sign any new revocations.
    \item If all the following conditions hold, the party sent a valid revocation to the block\-chain:
    \begin{enumerate}
        \item The channel traced by the party was in the \texttt{DisputeOpen} state;
        \item The party has a revocation to send;
        \item Messages that are sent now are guaranteed to be delivered to the block\-chain before the channel's deadlock is released;
    \end{enumerate}
    \item One of the following conditions holds:
    \begin{enumerate}
        \item The party started tracking a channel in a normal state (this case is used for channel opening);
        \item The party was tracking a channel in a normal state before and after the transition (this case is used during normal channel operation);
        \item \texttt{balanceOur} (see Item~\ref{it:balanceOur} in \texttt{goodSimpleParty})
              did not change during the transition (this case is used for channel closing).
    \end{enumerate}
\end{enumerate}

\end{document}